\documentstyle[psfig]{mn2e}

\def\etal{{\it et al.~}}

\def\half  	{\textstyle {1 \over 2} \displaystyle}

\def\inv16  	{\textstyle {1 \over 16} \displaystyle}
\def\3ov4  	{\textstyle {3 \over 4} \displaystyle}
\def\4ov3  	{\textstyle {4 \over 3} \displaystyle}
\def\8ov11  	{\textstyle {8 \over 11} \displaystyle}
\def\15ov16  	{\textstyle {15 \over 16} \displaystyle}

\def\beq{\begin{equation}} \def\eeq{\end{equation}}
\def\bea{\begin{eqnarray}} \def\eea{\end{eqnarray}}

\def\msun{	{\rm ~M}_\odot}
\def\AM{ 	\hat{\rm A} } 	
\def\qB{ 	\hat{q}_{_{\hat{\bf B}}} } 	
\def\hchi{ 	\hat{\chi} }

\def\sA{{\cal A}}
\def\sC{{\cal C}}
\def\sD{{\cal D}}

\def\cH{{\cal H}}
\def\cO{{\cal O}} 

\def\ccH{ c_{_\cH}}
\def\Valf{{\it v}_{_{\rm Alf}}}

\def\Oor{{\rm Oort}}
\def\AG{{\rm Araya-G\'ochez\,}}

\def\Prad{p_{\rm rad}}
\def\Pgas{p_{\rm gas}}

\def\PB{p_{_{\rm \bf B}}}

 	\def\bk{ {\bf k}}	
\def\bB{ {\bf B}}

\def\bxi { {\mbox{\boldmath $\xi$}} }
\def\unvc{ {\mbox{\boldmath $1$}} }

\def\Del { {\mbox{\boldmath $\nabla$}\!} }

\def\hom{\hat{\sigma}}

\def \ltaprx {\lower .1ex\hbox{\rlap{\raise .6ex\hbox{\hskip .3ex
	{\ifmmode{\scriptscriptstyle <}\else 
		{$\scriptscriptstyle <$}\fi}}}
	\kern -.4ex{\ifmmode{\scriptscriptstyle \sim}\else 
		{$\scriptscriptstyle\sim$}\fi}}}
\def\gtaprx {\lower .1ex\hbox{\rlap{\raise .6ex\hbox{\hskip .3ex
	{\ifmmode{\scriptscriptstyle >}\else 
		{$\scriptscriptstyle >$}\fi}}}
	\kern -.4ex{\ifmmode{\scriptscriptstyle \sim}\else 
		{$\scriptscriptstyle\sim$}\fi}}}

\def\sec{{\rm ~sec}}

\begin{document}


\title{Gravitational Waves from Hyper-Accretion onto Nascent Black Holes}

\author [R. A. Araya-G\'ochez]
	{Rafael Angel 
	Araya-G\'ochez \\
  	Theoretical Astrophysics MC 130-33,
  	California Institute of Technology, Pasadena CA 91125	
	\,\, {\bf arayag@tapir.caltech.edu}	
	} 

\onecolumn  	
\date{}
\pubyear{2002}
\maketitle

\begin{abstract} 
	We examine the possibility that hyper-accretion onto newly born,
 black holes occurs in highly intermittent, non-asymmetric fashion 
 favorable to gravitational wave emission in a neutrino cooled disk.
 This picture of near-hole accretion is motivated by 
 magneto-rotationally induced, ultra-relativistic disk dynamics in the region 
 of the flow bounded from below by the marginally bound geodesic radius 
 $r_{\rm mb}$.
 For high spin values,  
 a largely coherent magnetic field in this region has the dynamical 
 implication of compact mass segregation at the displacement nodes 
 of the non-axisymmetric, MRI modes.
 When neutrino stress competes favorably for the disk dynamical structure, 
 the matter clumps may be rather dense and sufficiently long-lived
 to excite the Quasi-Normal Ringing (a.k.a. QNR) modes of the Kerr
 geometry upon their in-fall.
 We find that such accretion flow may drive  
 bar-like, quadrupole $(l,m=2,2)$ modes in nearly resonant 
 fashion for spin parameters $a \geq .9$.
 The ensuing build up in strain amplitude of the undamped oscillations
 warrants a brisk rate of energy deposition into gravitational waves.
 A detectability assessment 
 for the LIGO interferometers through the match filtering technique
 is given by integrating the energy flux over a one second epoch of 
 resonant hyper-accretion at $1 \msun \sec^{-1}$.
 Thus, 
 a $15 \msun$ Kerr black hole spinning at $a \simeq .98$
 ($f_{\rm QNR} \simeq 1677$ Hz),
 and located at 27 Mpc (e.g., GRB980425), will deliver a
 characteristic strain amplitude, $h_{\rm char} \simeq 2.2_{-21}$,
 large enough to be detectable by LIGO II. 
 If resonant hyper-accretion were sustainable 
 for a longer period (or at higher rates) 
 possibly associated with a second broad hump in a GRB light-curve, 
 these objects could be detected by LIGO I at very low redshifts.
\end{abstract}    

\begin{keywords}
MHD---instabilities---black hole physics---gravitational waves
\end{keywords}

\section{Introduction}
 \label{sec:Intro}

 It seems only fitting for ultra-relativistic, black hole accretion 
 to be the leading contender to energize and explode the massive stars 
 associated with gamma-ray bursts and hyper-nov\ae.
 One must recognize, however, that such a scenario resides on a very exotic 
 front: stellar-mass black holes accreting at twelve orders of magnitude
 above the Eddington limit!
 On the other hand, 
 since GRB engines are effectively hidden from view, gravitational wave 
 emission models may constitute the most effective probes to such 
 enigmatic events. 
 Indeed, there is a great deal of impetus to address such 
 extreme accretion scenario and 
 its associated, non-standard energy deposition channels.
 In this paper we are chiefly concerned with the possibility
 that the dynamical structure of hyper-accreting flows is highly
 intermittent on large scales, with the in-fall of large mass 
 over-densities leading to prolific gravitational wave emission
 coincident with the gamma-ray stage. 

	We envisioned the accretion disk setting following the 
 hyper-accreting black hole models of Popham, Woosley \& Fryer (1999, 
 hereafter PWF).
 For a fiducial scenario: ${\rm M}_{h} = 3 \msun, ~\alpha_{_{\rm SS}} = 0.1,
 ~{\rm and}~ \dot{\rm M} = 0.1 \msun \sec^{-1}$, 
 these authors found that the onset of photo-disintegration 
 and of neutrino cooling at radii $r \ltaprx \,70 ~[GM/c^2]$ 
 yield a mildly advective,
 semi-thin, $\cH_\Theta / r \ltaprx .4$, accretion disk structure
 with a slightly sub-Keplerian rotation profile.  
 An improved account of neutrino transport (Di Matteo \etal 2002)
 has found that the innermost disk portion should become optically thick 
 to neutrinos for accretion rates 
 $\dot{\rm M} \gtaprx 0.1 \msun \sec^{-1}$.
 This being the case, cooling by radial advection of energy will compete
 with that from local neutrino emission and advection will 
 overwhelm neutrino cooling for $\dot{\rm M} \gtaprx 1 \msun \sec^{-1}$
 (Di Matteo \etal 2002). 
 Yet, such a high accretion rate is rather unlikely 
 (MacFadyen \etal 2001), so we restrict our analysis to 
 $\dot{\rm M} \leq \,1 \msun \sec^{-1}$

	The structure of this paper is a follows.
 	In \S 2, we assess the relevance of physical processes occurring 
 in the relativistic region of the flow: $r_{\rm mb} \leq r \leq r_{\rm ms}$.
 Magneto-rotationally-induced, relativistic disk dynamics is addressed in 
 \S 3, where we also discuss the role of compressibility and estimate
 the size of mass over-densities from a neutrino cooled disk. 
 In \S4 we build an idealized, analytical model for gravitational wave 
 emission and, in \S 5, estimate the detectability of the GW signal
 for the LIGO interferometers with the match filtering technique.

\section{Physics Inside the ISCO}
\label{sec:ISCOphys}

	The model for gravitational wave emission 
 outlined in \S \ref{sec:Model} is based on the premise
 that large-scale magneto-rotational effects will drive the disk 
 dynamics in the ultra-relativistic region of the flow bounded from 
 below by the marginally bound orbit radius. 
 For spin rates $a \geq .9$, the growth rate of the magneto-rotational 
 instability $\sigma_{\rm MRI}$ (a.k.a. MRI),
 is faster than the inverse dynamical time scale, $\Omega_+^{-1}$,
 throughout the annulus 
 $r_{\rm mb} \leq r \leq r_{\rm ms}$
 by a factor $\simeq \cO[2]$ (see Fig 1). 
 Thus, a small delay to reach the ``free-fall" stage is all that is 
 needed for magneto-rotational dynamics to take place.
 On the other hand, 
 $r_{\rm mb}$ represents the lowest bound on the location of the cusp 
 in effective potential and is akin to a relativistic generalization 
 of the inner Lagrangian point, L$_1$, for mass transfer in close binaries 
 (Kozlowski, Jaroszy\'nski \& Abramowicz 1978).

	The exact location of the inner boundary depends on the source 
 of free-energy for the flow.
 Standard lore holds that the flow draws energy from the radial 
 gradient in angular momentum  
 and that since dissipation of angular momentum vanishes
 at $r_{\rm ms}$ for a Keplerian disk, this is where the inner edge of 
 such a disk should reside (Novikov \& Thorne 1973).
 Yet, we now know that when the disk dynamics is driven by magneto-rotational 
 effects the source of free energy is not the gradient in the radial 
 distribution of angular momentum but rather the gradient in angular 
 velocity\footnote{More generally, 
Balbus (2002) has found that for a magneto-rotationally-driven disk,
the proper replacement of the classical H{\o}iland stability criteria 
for rotational and convective motions involves the gradients 
in angular velocity and temperature instead of angular momentum and entropy.
}; that is, the source of free-energy is 
 the shear of the congruence of circular geodesics. 
 Thus, MRI-mediated, turbulent angular momentum transport goes on 
 unabated at the radius of marginal stability for cold geodesic orbits 
(\AG 2002) 
 and the flow within need not preserve specific angular momentum
 {\em nor} loose its relativistic Keplerian rotation profile
 in spite of a non-trivial radial velocity.  

	We will therefore assume a Keplerian angular velocity profile 
 throughout: $r \geq r_{\rm mb}$.  
 This assumption is consistent with the models of PWF and 
 Popham \& Gammie (1999)--indeed, strong deviations from Keplerian rotation 
 only occur as fluid elements approach the photon radius $r_{\rm ph}$--and 
 it is also consistent with global 3D simulations of non-radiative 
 MHD accretion onto non-spinning black holes 
 (Hawley \& Krolik 2001, Hawley \& Balbus 2002)
 in spite of partial pressure support for hot inner tori.  

 	Inside $r_{\rm ms}$, the flow may not have time to cool significantly 
 and advection of entropy will become progressively more important as the 
 horizon, $r_+$, is approached.
 Entirely advective, accretion flows 
 (i.e., non-radiative, hydrodynamical flows) 
 are reckoned to posses a large, positive Bernoulli function
 ${\cal B} = \varrho + \int_0^p \! \! dp \, \rho/\varrho$
 (Abramowicz \etal 1978, Blandford \& Begelman 2003) 
 which makes the ultimate fate of accreting fluid particles 
 a theme of controversy;
 e.g., ADAF (Narayan \& Yi 1995) {\it vs} 
 CDAF (Narayan \etal 2000) {\it vs} 
 ADIOS (Blandford \& Begelman 1999, 2003). 
 The broad conclusion drawn from this ongoing debate is that 
 fluids with large internal energies, e.g., strongly magnetized fluids, 
 are not easily accreted onto gravitational wells.
 In the ultra-relativistic flows of interest to us,
 and for large spin parameter $a$, 
 the value of the relativistic enthalpy, 
 $\varrho \equiv (\rho + \varepsilon + p)$, can be non-trivial 
 $\varrho/\rho \geq 2$
 (with magnetic energy folded into the internal energy $\varepsilon$),
 implying a high likelihood for large Bernoulli function as well.
 We adopt the view that for a fluid with such high enthalpy, 
 the dynamical boundaries set by the circular orbits of {\it cold} 
 (point) particles in the Kerr geometry are inadequate.

\section{Magneto-rotationally-induced Relativistic Disk Dynamics}

	The MRI enables accretion through vigorous, turbulent transport 
 of angular momentum.  
 It also constitutes the key process to tap the free-energy available 
 in a flow endowed with differential rotation.
 In the Newtonian picture, this is essentially a local, co-moving instability
 as the interplay of inertial accelerations with the elastic coupling of 
 fluid elements creates an unstable situation 
 to the redistribution of specific angular momentum, $\ell(r)$, if
 the {\it angular velocity profile} decreases monotonically with radius.
 This criterion is in opposition to the (non-magnetic) 
 Rayleigh criterion for stability of a differentially rotating fluid: 
 $r^{-3}$ d$_r \ell^2 = \kappa^2 \geq 0$, 
 where $\kappa$ is the frequency of radial epicyclic motions. 
 Indeed, 
 without the elastic coupling provided by the bending of field lines, 
 such inertial forces--namely, the shear(tide) and the coriolis terms--induce
 radial epicyclic motions while preserving specific angular momentum. 
 
	Incompressible
 MRI initiated turbulence has a characteristic ``parallel" scale 
 for fastest growing modes corresponding to 
 $k_\parallel \equiv \bk \cdot \unvc_{\bB} \simeq \Omega/\Valf$,
 and a normalized growth rate 
 $\sigma = \AM \equiv \half {\rm d}_{\ln r} \ln \Omega$, 
 associated with the shear parameter, a.k.a. the Oort A ``constant".
 Although slower growing modes occur at larger scales, we take 
 the fastest growing modes to be the dynamically predominant eddies 
 given that the radial flow velocity is non-negligible 
 in $r_{\rm mb} \leq r \leq r_{\rm ms}$. 
 Thus, the scale associated with the MRI is generally
 smaller than the vertical scale height of the disk by a factor 
 $\cO [\Valf/c_\cH]$ and no large-scale effects are expected.
 Indeed, in the weak field limit one may construct a dispersion relation 
 quite independently of the large-scale field topology: 
 highly sub-thermal fields, $\Valf/c_\cH \ll 1$, guarantee that  
 the instability is truly local and incompressible (see below).
 When the field is more moderate,
 $k_\parallel^{-1}$ will approach the disk's pressure scale height
 $\cH \simeq c_\cH / \Omega$. 
 and this imposes a lower bound on $k_\parallel$
 when the field is mostly vertical.  
 On the other hand, the lowest bound on $k_\parallel$
 in a toroidal field geometry is $1/r$ which would correspond 
 to a supra-thermal field in the Newtonian picture
 (Foglizzo \& Tagger 1995, hereafter FT).
 Because of the possibility of large-scale, global disk dynamics, 
 these modes are the focus of our analysis below.

\subsection{Compressible Non-asymmetric MRI modes}
\label{subs:Compress}

 	In the 2D regime of fastest growth,
 $k_\theta \gg k_r \gg k_\varphi$, 
 non-axisymmetric ``horizontal" displacement modes of a toroidal field 
 have the following dispersion relation (\AG 2002)
\beq 		\hom^4 
	- \{ (\Lambda + 1) \qB^2 + \hchi^2 \} \, \hom^2
	+ 	\Lambda \, \qB^2 \{  \qB^2 +  4 \AM \}
	= \emptyset,
\label{eq:disper} \eeq
 where all frequencies  
 are normalized to the rotation rate, 
 $\AM \equiv \half {\rm d}_{\ln r} \ln \Omega$ is the Oort A ``constant",
 $\hchi^2 \equiv 4(1+\AM)$ is the squared of the epicyclic frequency, and 
 $\qB \equiv ({\bk} \cdot {\bf v}_{\rm Alf}) / \Omega$ 
 is a frequency related to the component of the wave vector along the 
 field (in velocity units). 

 $\Lambda$ is defined through
\beq 	
	\Lambda  
	\equiv {\Gamma \over {\Gamma + \Theta}} 
\label{eq:Lambd} \eeq
 where $\Gamma$ is the adiabatic index,
 and $\Theta \equiv \Valf^2 / c^2_\cH$.  	
 This definition entrains an important anisotropy constraint 
 on the Lagrangian displacement vector field 
 $\bxi: 
 \Valf^2 \, (\bk_\perp \cdot \bxi_\perp) = -c_s^2 \, (\bk \cdot \bxi)$,
 as well as the compressibility characteristics of non-axisymmetric modes
 (as a function of the component of Lagrangian displacement along the field,
 Foglizzo and Tagger 1995):
\beq{\Delta \rho \over \rho} = (1 - \Lambda)~ (-i k_\parallel \xi_\parallel)  
\label{eq:Compress} \eeq
 where $\Delta$ denotes the Lagrangian (co-moving) perturbation 
 as opposed to the Eulerian perturbation $\delta$
 (recall the non-relativistic relation 
 $\tilde{\Delta} = \delta + \bxi \cdot \Del$). 
 Evidently, the compressibility of the modes is 
 imprint on the deviations of $\Lambda$ from unity.
 A word of caution here regards the peculiar behavior of the fluid 
 when the internal stress has heat conduction characteristics, 
 e.g. photon or neutrino pressure; see \S \ref{subs:Heat&Clump}.  
 From Eq [\ref{eq:Compress}], one reads that the degree of 
 compression of the modes gets stronger with the field strength 
 and, naturally, with a softer equation of state.
 Note that setting $\Lambda \doteq 1$ in the dispersion relation 
 Eq [\ref{eq:disper}] yields the incompressible, local variant 
 of the MRI. 

	From the dispersion relation Eq [\ref{eq:disper}], 
 one finds the parallel wave-numbers of fastest growth to satisfy 
 (see also Foglizzo 1995)
\beq
	\qB^2 =	- 2 \AM 
	+ ( {\textstyle {{1 + \Lambda}\over{2\Lambda}} ) 	
	\times \left\{ -{2 \Lambda \AM^2 \over D} \right\}\displaystyle}
 	~~\stackrel{\Lambda \rightarrow 1}{\longrightarrow}~
	-\AM (2+\AM),
~~{\rm where}~~
	D \equiv 1 
		+  ({\textstyle { {1 - \Lambda}\over2 }\displaystyle}) \AM	
		+  \sqrt{1 + (1 - \Lambda) \AM} ,	
\label{eq:qB-om} \eeq
 while the expression in the curly brackets corresponds to 
 the square of the growth rate.

\subsection{General Relativistic Effects in the Cowling Limit}

	General relativity modifies this Newtonian results in that 
 it introduces large-scale effects when shear begins to overwhelm
 the coriolis terms, i.e., as $r_{\rm ph}$ is approached
 (see Eqs [32] of \AG 2002). 
 Although the hole is likely to be born in an excited state, 
 e.g., away from its asymptotic, Kerr geometry;
 we work in the Cowling limit, $\delta g \doteq \emptyset$,
 and, to zeroth order, use the standard form of the Kerr metric
 in the equatorial plane (Boyer-Lindquist coordinates):
\beq
ds^2 = -{\sD\over{\sA}} dt^2 + r^2\sA(d\varphi - \omega d t)^2
        + {1\over{\sD}}dr^2,
\label{eq:Metric} \eeq
 with $\omega \equiv {2 a / {\sA r^3}}$
 the rate of frame dragging by the hole 
 and where the metric functions of the radial {\bf BLF} coordinate are written
 as relativistic corrections (e.g. Novikov and Thorne 1973):
\[
\sA \equiv 1 + {a^2\over r^2} + 2 {a^2\over r^3},
~~{\rm and}~~
\sD \equiv 1 - {2\over r} + {a^2\over r^2},
\]
 in normalized geometrical units ($c = G = M_{_{\rm bh}} = 1$).

 	For circular, incompressible, geodesic flow, 
 the dispersion relation near a rotating hole is given by (\AG 2002)
\beq
	(\gamma\sigma)^4
	- 
	\left[ 
 	2  q_{\hat{\bB}}^2 + 
	{4 \gamma^2} \Omega_\pm^2 
	\left( \sC_\pm - \3ov4 \sD \right) 
	\right] (\gamma\sigma)^2 
	+
	q_{\hat{\bB}}^2 \,
	\left[ q_{\hat{\bB}}^2 - {4} \, 
 	\left\{\3ov4 \gamma^2 \sD \, \Omega^2_\pm \right\} \right]
	= \emptyset
\eeq
 where $\gamma$ is the red-shift factor, $\sD = g^{-1}_{rr}$,
 and 
 $\sC_\pm \equiv {1 - {3\over r} \pm { 2a \over r^{3/2}} }$
 corresponds to the $\sC$ function of Novikov \& Thorne (1973)
 for pro-grade (+) orbits.
 Regarding the electromagnetic stress,  
 the only difference with the non-relativistic, incompressible analog is that 
 the Alfv\'en speed is now weighted by the relativistic enthalpy of the fluid
 (Araya-Gochez 2002)
\beq 	\varrho \Valf^2 \equiv \half F^{\mu\nu} F_{\mu\nu}.
\label{eq:RelAlf}\eeq

	Factoring out the orbital frequency
 $u^\varphi / u^t \equiv \Omega_\pm = \pm (r^{3/2} \pm a)^{-1}$ and with  
 $\gamma \sigma \equiv \Omega_\pm \, \hat{\sigma}$, one has 
\beq
	\hat{\sigma}^4 
	- 
	\left[ 2 \qB^2 + \hchi_\pm^2 
	\right] \hat{\sigma}^2 
	+
	\qB^2 \left[ \qB^2 +  4 \AM \right]
	= \emptyset
\label{eq:GenRelDisp} \eeq
 where 
\[ \AM \equiv - \left\{\3ov4 \gamma^2 \sD \right\}
~~{\rm and}~~
\hchi^2_\pm = 4 \gamma^2 \left( \sC_\pm - \3ov4 \sD \right)
\]
 denote the normalized shear parameter and co-moving epicycle frequency 
 (${1 \over \gamma} \, \hchi$ corresponds to 
 the epicycle frequency as measured at asymptotic infinity).  
 Thus, 
 with the proper generalizations of the epicycle frequency and shear 
 parameter, the local dispersion relation is identical with 
 the Newtonian case in the limit of no fluid compression,
 c.f.  Eqs [\ref{eq:disper}\&\ref{eq:Compress}] 
 with $\Lambda \rightarrow 1$.

	Next, 
 using the relation $\gamma^2 = (1 \pm a/r^{3/2})^2 \, \sC_\pm^{-1}$
 for circular, geodesic flow (Novikov and Thorne 1973), 
 one finds the fastest growing modes will conform with
\beq
	\qB^2 	= 1 - \inv16 \hchi^4 
		= 1 - 
		\left( 1 \pm {a \over {r^{3\over2}}} \right)^4 \,
		\left\{ 1 - {3\over4} \, {\sD \over \sC_\pm} \right\}^2
\label{eq:WaveNumb} \eeq 
 which remains finite and close to the Newtonian value 
 $\qB \rightarrow -\AM (2 + \AM)  = 15/16$ 
 for all radii outside of the ISCO.
	
	Recall that the marginally stable orbit (a.k.a. the ISCO),
 $r_{\rm ms}$, corresponds to the root of $\hchi_\pm = 0$. 
 The radius of the circular photon orbit, $r_{\rm ph}$,
 is where $\sC_\pm = 0$ and the event horizon, 
 $r_+$, happens at the outer root of $\sD = 0$. 
 For any value of the rotation parameter 
 $a:~ r_{\rm ms} > r_{\rm mb} > r_{\rm ph} > r_+$.
 Our statement on large-scale effects follows from these remarks 
 and from direct inspection of Eq [\ref{eq:WaveNumb}]: 
 $\qB \rightarrow 0^+$ as $r \rightarrow r^+_{\rm ph}$, 
 i.e. the most unstable MRI modes are go to large scales 
 as the photon orbit is approached.
 Figs 1 \& 2 show the general trends for the normalized growth rate 
 and wave-numbers as functions of radius and spin parameter.

\begin{figure}
\hbox{~}
\centerline{\psfig{file=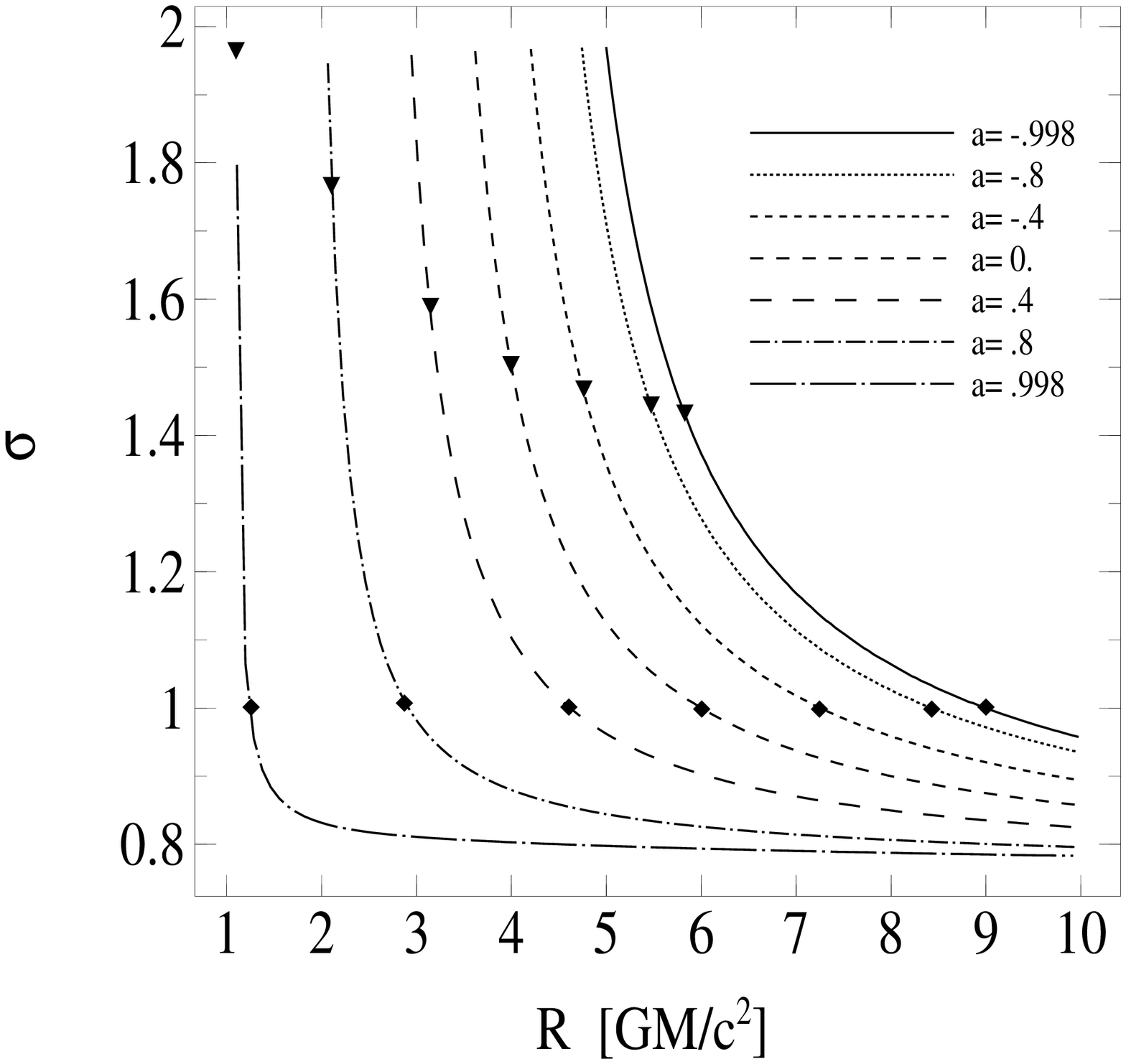,angle=0,width=5truein}}
\vskip 1mm 
\caption{Normalized growth rate, $\hat{\sigma}$, as a function of radius
for several values of the spin parameter $a$ (from \AG 2002).
}
\end{figure}

	Two notable caveats of the relativistic approach involve 
 the global disk structure and the compressibility of the modes.
 We discuss these in turn.  

 Curvature effects and radial field structure 
 will indeed modify the dispersion 
relation\footnote{
the interested reader is invited to compare 
the relativistic, local version of the equations of motion for $\xi$:
Eqs [32] of \AG 2002, with the Newtonian, global version 
Eqs [2.6 \& 2.7] of Ogilvie \& Pringle 1996
}
 (see, e.g., Curry \& Pudritz 1995, Ogilvie \& Pringle 1996)
 but these effects are unlikely to modify 
 the gross properties of the fastest growing modes unless
 radial stratification plays a destabilizing role. 
 If such stratification were magnetically-driven;  
 it would require an unlikely supra-thermal field 
 $\Valf \approx v^{\tilde{\varphi}} = \tilde{r} (\Omega - \omega)$,
 where $\tilde{r} \equiv r \sA / \sqrt{\sD}$ 
 is the radius of gyration for the physical velocity in 
 the locally non-rotating frame (Bardeen, Press \& Teukolsky 1972).
 Furthermore,
 we note here and {\em correct} our previous comment (\AG 2002):
 $\qB \rightarrow 0^+$ is merely a statement on the 
 length-scale of the fastest growing modes, 
 $k_\parallel^{-1}\simeq \cO [ \tilde{r}]$, 
 regardless of field strength. 

 	The compressibility of the modes
 alters the threshold of shear parameter where $\qB \rightarrow 0^+$,
 but does not fundamentally affect the conclusions drawn from  
 Eq [\ref{eq:WaveNumb}] either. 
 For roughly equal neutrino and radiation pressures (see below), 
 one can use an effective adiabatic index 
 (\AG \& Vishniac 2002) in Eq [\ref{eq:qB-om}]
 to anticipate that the effects of compressibility on toroidal modes
 is to increment the threshold of shear parameter 
 where $\qB \rightarrow 0^+$ 
 from $-\AM = 2$ to $-\AM \rightarrow 1 + 2 D / (1 + \Lambda)$;
 c.f. Eq [\ref{eq:qB-om}].
 Nevertheless, since for geodesic flow $\AM \propto \sD/\sC_\pm$ and 
 $\sC_\pm \rightarrow 0$ @ $r_{\rm ph}$, 
 the increase in shear threshold in this setting is rather inconsequential.

\begin{figure}
\hbox{~}
\centerline{\psfig{file=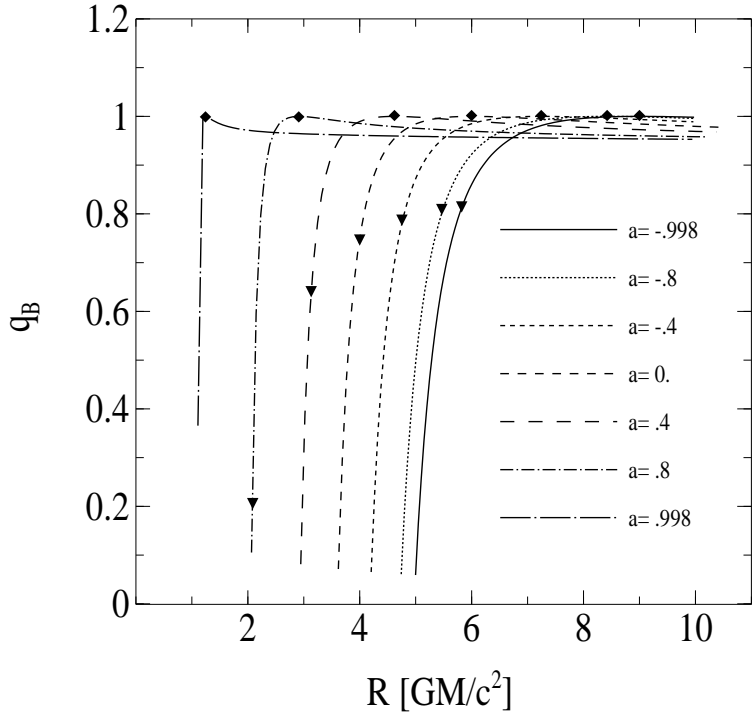,angle=0,width=5truein}}
\vskip 1mm 
\caption{Normalized parallel wavenumber, $\qB$, as a function of radius
(in gravitational radii)
for several values of the spin parameter $a$.
Diamonds indicate the location of the marginally stable orbit,
$\hchi \doteq \emptyset$,
and triangles, the location of the marginally bound orbit (from \AG 2002).
}
\end{figure}

\subsection{Heat Conduction and Clump Scale}
 \label{subs:Heat&Clump}

	A precise assessment of radiative diffusion effects
 in the general relativistic regime is hampered by the breakdown of a 
 key assumption made to simplify the ``linear poking" on the Faraday 
 field tensor (\AG 2002):
 use of the enthalpy weighted specific four magnetic field 
 in Eq [\ref{eq:RelAlf}]. 
 However, 
 just as in radiation pressure dominated fluids,
 one expects a neutrino stress dominated 
 gas--with neutrinos semi-contained by the nucleon component 
 through scattering--to comprise a rather peculiar MHD fluid. 
 Indeed, when a fluid transitions into a radiative pressure regime, 
 compressive modes will loose pressure support in a given range of 
 wave-numbers (Agol \& Krolik 1998). 
 Thus, 
 the magnetic field is truly frozen only to the co-moving volume associated 
 with the baryon component and compressive perturbations need not 
 behave adiabatically.

 	The half optical depth to scattering of a 
 standard radiation-pressure--dominated $\alpha$-disk 
 is related to the scale height, $\cH$, and rotation rate, $\Omega$, by
 $\tau_{\rm disk} =  - c / (2 \alpha \AM \Omega \cH)$,
 This relation is explicitly sensitive only to the local nature of the 
 cooling but is implicitly subject to a suitable vertical gradient 
 of heat deposition (see, e.g., Krolik 1999).
 Therefore, these scalings also hold for a neutrino cooled $\alpha$-disk
 through a replacement in the source of opacity by neutrino scattering
 in the non-advective accretion regime:
 $0.1 \msun \sec^{-1} \leq \dot{\rm M} \leq 1 \msun \sec^{-1}$. 
 On the other hand, 
 in the Newtonian regime $\qB \simeq \cO [1]$  
 and for magnetic angular momentum transport such that  
 $\alpha \simeq \Valf^2 / \ccH^2$,
 the depth through the MRI eddies is 
 $\tau_{\rm eddy} \sim {\Valf/\ccH} \times \tau_{\rm disk}
	\sim c / (2 \AM \sqrt{\alpha} \ccH )  \gg 1$. 
 Thus, assuming an isotropic random walk for neutrinos,
 the diffusion time through these eddies, 
 $t_{\rm diff} \sim \Omega^{-1}$, 
 is similar to the time scale for fastest MRI modes to develop
 $t_{\rm MRI} \simeq -{\rm A}^{-1}_\Oor$.

	Radiative heat conduction will isotropize the dynamically dominant 
 modes since neutrinos will diffuse through the smallest distance 
 associated with the eddies, i.e. mostly in the vertical direction.   
 In our problem, 
 this will diminish the growth rate of the instability
 (Blaes \& Socrates 2001) but not by much. 
 Indeed, if neutrino stress were predominant, 
 the horizontal regime of fastest growth (e.g. \S \ref{subs:Compress}) 
 would be inaccessible;  
 yet, such is never the case since the pressure contributions 
 from radiation, pairs and neutrinos, 
 all have the same temperature dependence $a/6~T^4$
 with the relative contributions 
 varying only by internal degrees of freedom 
 times particle statistics factors  
 (with only one helicity state for neutrino pairs):
 2 $\times $ 1, 4 $\times$ 7/8, and 6 $\times$ 7/8 respectively.  
 Thus, $p_\nu$ is never greater than about $\simeq \Prad$
 where $\Prad = 11/12 a T^4$ includes the pressure from 
 pairs and photons.

	Let's quantify these arguments in order to gauge the 
 length scale of the clumps.  At a very fundamental level,
 clump formation is intimately connected to the effects of 
 radiative heat conduction out of compressive perturbations.
 This can be understood by writing down the polarization properties 
 of the fastest growing modes in the horizontal regime:
 $\xi_r = - \sqrt{\Lambda} ~\xi_\varphi~\&~|\xi_\theta| \ll |\xi_r|$,
 and reading from Eq [\ref{eq:Compress}] that 
 at the linear stage of the instability there exists 
 a converging flow toward the Lagrangian displacement node of the modes
 (see Fig 2 of Foglizzo and Tagger 1995). 
 In a fluid with entirely elastic (adiabatic) properties, 
 the pressure perturbation associated with such compression will 
 act as a restoring force to de-compress the fluid in the non-linear stage.   
 On the other hand, 
 when the fluid has radiative heat conduction properties 
 on the scale of the density perturbations, 
 no such restoring force persists on time scales longer than 
 the inverse of the growth rate 
 so the clumps will survive for longer times.
 Turner \etal (2001, 2003) report that in a standard 
 radiation-pressure--dominated disk, 
 when $\Prad \gtaprx \Pgas \simeq \PB$,
 the non-linear outcome of the MRI is a porous medium
 with drastic density contrasts.
 Under nearly constant total pressure and temperature, 
 the non-linear regime shows that density enhancements 
 anti-correlate with azimuthal field 
 domains--just as expected from the linear theory--and 
 that turbulent eddies live for about a dynamical time scale
 while mass clumps are destroyed through collisions or by running 
 through localized regions of shear on a similar time scale.
 Notably, the non-linear density contrasts may be quite large
 $<\rho_{\rm max}/\rho_{\rm min}> \gtaprx~ \cO[10]$.   
 We make no attempt here to estimate 
 $\Delta \rho / \rho$ from the linear theory
 but rather take the view that when the length scale of the 
 instability is suitable large, 
 $\lambda_{\rm MRI} \simeq  \cO [\tilde{r}]$ 
 (see below), a fraction of $\cO[1/2\pi]$ of the mass in the annulus 
 $r_{\rm mb} \leq r \leq r_{\rm ms}$ will 
 reside in one massive clump such that the gravitational wave emission 
 associated with the in-fall of this clump will dominate the power spectrum.

	Let us emphasize that the isotropic size of the clumps is determined 
 by the condition that, on the time scale of the instability, neutrinos 
 will diffuse through the clumps {\it via} nucleon scattering 
 and/or absorption followed by re-emission.
 For simplicity, we again assume local balance of dissipation from 
 the shear flow with cooling from neutrino emission and 
 a constant supersonic Mach number for the disk flow:
 $M_\varphi = v^{\tilde{\varphi}} /c_\cH$
 ($M_\varphi = 2.5$ is consistent with the results of PWF in the 
 neutrino cooling regime). 
 Under this assumption, 
 we need not worry about detailed neutrino transport 
 and, relaxing the condition $\qB \approx 1$, 
 the size of the most massive clumps will correspond to 
\[ 	l_{\rm clump} 	\simeq \Valf/c_\cH~\cH 
			\sim \sqrt{\alpha} M^{-1}_\varphi \tilde{r}.
\] 
 Thus, in the relativistic regime
 the length-scale of the clumps will be smaller 
 than the scale associated with the fastest growing modes by a factor
 of $\cO [\qB]$.

\section{A Simple Physical Model for Gravitational Wave Emission}
\label{sec:Model}


	The key to our argument for clumpy, near-hole hyper-accretion
 is based on the observation that the length-scale of the fastest 
 growing modes may be quite large at the innermost disk boundary
 $r_{\rm mb}$.
 Since the radial velocity is non-negligible, one cannot demand 
 ``tightly fitting" eigenmodes on the circumference at fixed radius. 
 Loosely speaking, the WKB modes live on ingoing spiral trajectories
 (see \S \ref{subs:ResDrivQNR});
 but we are only concerned with their overall gross properties.

 	The clumps form very rapidly, 
 $t^{-1}_{\rm MRI} \simeq \Omega\hat{\sigma}$, and very close to the horizon. 
 Since these are very compact, $l/\tilde{r} \simeq \cO [0.1]$,
 we considered them as point particles in the order of magnitude estimate
 of gravitational wave emission below.
 Computation of accurate waveforms for the in-fall of a single clump
 constitutes a daunting task because of the 
 non-trivial multipole contributions 
 to the radiation field in the strong gravity regime near $r_{\rm mb}$.  
 The contributions from higher than quadrupole will broaden the power 
 spectrum of emitted waves 
 (Zerilli 1970, Davis \etal 1971, hereafter DRPP) 
 and recent results by Lousto \& Price (1997, hereafter LP97)
 show that particles released from very near the marginally bound orbit
 of a non-spinning hole--or 
 more precisely near the maximum of the Zerilli potential--will 
 maximize the locally radiated energy.

	As a ``back of the envelope" estimate of the quadrupole r.m.s. strain,
 $|h| = \sqrt{h^2_+ + h^2_\times}$, associated with the in-fall of a single 
 clump, we simply assume that the clump is formed above $r_{\rm mb}$
 and that it survives for one rotation period before plunging into the hole.
 The transverse-traceless projection of the metric perturbation 
 in the radiation zone, $h^{\rm TT}_{ij}$, is estimated by replacing 
 the second time derivative of the mass quadrupole moment,
 $\partial_t^2 {\cal J}^{\rm TT}_{ij}$, 
 by the kinetic energy associated with non-spherical motion; 
 i.e., clump rotation as seen at asymptotic infinity 
 (Thorne 1987). Thus
\[ 	|h| = |h^{\rm TT}_{ij}| \equiv 
		{2 \over d} \left[ G \over c^4 \right] 
		\left| 
		\partial_t^2 {\cal J}^{\rm TT}_{ij} (t-r) 
 		\right|
		\simeq \left[ G \over c^4 \right] 
		\frac{\tilde{r}^2 \Omega^2 }{d} \delta M  
		\simeq
		{G \delta M \over (c^2d)}.
\]
 
	For a $15 \msun$ black hole rotating at $a=.98$,
 and accreting $1 \msun \sec^{-1}$,
 the mass in the clumps is 
 $\delta M \simeq \dot{\rm M} / \Omega \simeq 1.8_{-4} \msun$,
 where $\Omega(r_{\rm mb}) \simeq .4 = 5400$ Hz; corresponding to 
 a linear frequency $\Omega(r_{\rm mb})/\pi = 1720$ Hz.
 This gives $|h| \simeq 3.2_{-25}$ for a source at 27 Mpc.
 To bring this into the LIGO band, 
 $|h|_{\rm eff} \simeq 1._{-21}$
 one would need to integrate over N $\simeq 1_{+8}$ cycles or
 over 1700 sec.! 
 Thus, the r.m.s. strain associated with in-fall of a single 
 clump onto the hole is too small to be of any astrophysical importance
 at present. 

 	But this issue is more subtle than it appears at first sight.

\subsection{Quasi-normal Ringing Waveforms: Single Excitation Event}
\label{subs:SingleQNR}

	Since the clumps form in a strong gravity regime, 
 the process of gravitational wave emission from in-fall at such 
 a short range must be cast as an excitation of the infinite number
 of increasingly damped quasi-normal modes of oscillation of the 
 background geometry (Leaver 1985). 
 The often-drawn analogy for such quasi-normal ringing (QNR) of the hole 
 to the ringing of a bell is fundamentally flawed in at least one 
 respect: the black hole is not excited by the smashing of the clump 
 as it ``hits" the horizon.  
 The hole is rather excited when the metric perturbation associated 
 with the clump is ``felt" by the background metric.
 The excitation event therefore constitutes a smooth process whereby 
 in-fall of a clump from $\simeq r_{\rm mb}$ and through $r_+$ 
 serves as a source in the Teukolsky (1973) equation 
 for small perturbations to the Kerr geometry
 (with appropriate boundary conditions at $r_+$ and $r_\infty$).
 This is an important distinction with a great deal of relevance 
 to the problem at hand since we need to gauge the ``driving" 
 of QNR modes in terms of an effective coupling from 
 clump in-fall.

 	Clumpy black hole accretion from an exterior disk will excite 
 preferentially quadrupole, bar-like mode perturbations of the geometry 
 with spheroidal harmonic
indices\footnote{
Quadrupole perturbations, $l=2$, are reckoned to contribute about 90\% 
of the energy budget for outgoing GW's while $l=4$ modes add about 10\%.
Higher multipole contributions are negligible,
see, e.g. DRPP, LP97.
} $(l, m) = (2, 2)$.
 An approximate analytical expression for the QNR frequency 
 of this mode is given by Echeverria (1989):
\beq
	\omega_{22} = \left[ 1 - 0.63(1-a)^{3/10} \right] 
	\times \left\{ 1 + {i \over 4}(1 - a)^{9/20} \right\}.
\eeq
 The corresponding quality factor,
 a.k.a. Q-value, of the (2,2) mode
\beq
 Q_{22}(a) \equiv {1 \over 2} {\Re [\omega_{22}] \over {\Im [\omega_{22}]}}
 		= 2 (1-a)^{-9/20}, 
\label{eq:Qval} \eeq
 is $\geq 2\pi$ for $a_{\rm crit} \geq 0.92$ 
 This represents a minimum value of $a$ to regard the hole as
 a decent bar-like--mode oscillator.
 Below $a_{\rm crit}$ the mode is damped too strongly to allow for any 
 significant build up of the strain signal/spectrum
 from resonant accretion (see below).       

	In the radiation zone and time domain, undriven QNR waveforms 
 constitute circularly polarized, damped oscillations 
 (Leaver 1985, Echeverria 1989) 	
\beq    h(t) \equiv h_+ - ih_\times
 	= {H_0 \over d} S_{22} (\phi,\theta,a) 
	~ e^{-i (\omega_{22} t - \varphi)}
\label{eq:DampQNR} \eeq
 where $d$ is the luminosity distance to the source,
 $S_{22} (\phi,\theta,a)$ is a spin weighted, 
 normalized ($\int d\Omega |S_{22}|^2 = 1$),
 spheroidal harmonic of coordinates $\phi$ \& $\theta$
 [in the Schwarzschild limit 
 $S_{22} (\phi,\theta,a \rightarrow 0) \longrightarrow
 Y_{22}(\phi,\theta)$].
 The strain length amplitude of the waveform, $H_0$ 
 (in units of the hole's mass M), and the phase shift, $\varphi$, 
 depend on the initial conditions of the metric perturbation.

 	$H_0$ gauges the efficiency of the emission process to posit
 a fraction of the total rest-mass energy of the system into 
 outgoing gravitational wave energy.
 For strong excitation events such as binary black hole coalescence, 
 this fraction may be as large as 3\% (see, e.g. Flanagan \& Huges 1998). 
 For the in-fall of {\it a} clump of mass $\delta M \ll M$, 
 this fraction scales as $\varepsilon (\delta M)^2/M$ (DRPP), 
 with a corresponding time domain strain amplitude 
 $\propto \sqrt{\varepsilon} \, \delta M / \sqrt{M}$. 
 The precise value of $\varepsilon$ for a rotating hole 
 is unknown, but it hovers on a few per cent for the perturbations of
 a Schwarzschild geometry induced by radial in-fall from 
infinity\footnote{
Strictly speaking, the energy released in this case includes 
not just the ringing phase but also a small contribution from 
the gravitational Bremsstrahlung component, 
see, e.g., LP97.   
} (DRPP).  
 For non-axisymmetric perturbations, the energy released from 
 clump in-fall may be considerably larger (Fryer, Holtz \& Hughes 2002) 
 and for nearly resonant, driven oscillations from hyper-accretion,   
 we demonstrate below that the total amount of energy deposited into 
 outgoing gravitational waves is only bounded from above by a factor
 of a few times the rest-mass--energy of a single clump!

\subsection{Collective Effects: Resonant Driving of QNR modes}
\label{subs:ResDrivQNR}

 To assess the free-fall stage of the clumps,
 we recall from \S \ref{sec:ISCOphys} that the cusp in effective potential 
 occurs somewhere above $r_{\rm mb}$ 
 (depending on angular velocity profile)
 and that a relativistically hot
 accretion flow may posses a semi-Keplerian angular velocity profile, 
 $\Omega_+ = (r^{3/2} + a)^{-1}$, down to this region,
 (PWF, Popham \& Gammie 1998, Hawley \& Krolik 2001).
 Parameterizing the accretion flow through its Mach number and 
 the angular momentum transport through a 
 Shakura \& Sunyaev ``$\alpha$" parameter (as in \S \ref{subs:Heat&Clump}, 
 we adopt M$_\varphi \doteq 2.5$ \& $\alpha \doteq 0.1$), 
 when $a \gtaprx .9$
 the wavelength associated with the fastest growing MRI modes becomes 
 comparable with $2 \pi \tilde{r}$ at a radius $r \geq r_{\rm mb}$; 
 e.g., when 
 $\qB \simeq .126 \sqrt{\alpha_{0.1}} \, {\rm M}^{-1}_{\varphi,2.5}$.
 At higher spin parameter values, MRI modes go to comparably 
 large scales at radii closer to $r_{\rm ms}$.
 Thus, the annulus $r_{\rm ms} \geq r \geq r_{\rm mb}$ constitutes a
 relativistic locus for large-scale magneto-rotationally induced fluid
 dynamics at moderately large values of black hole spin: $a \geq 0.9$.

	Since the radial velocity is non-negligible, MRI wave-modes 
 must be interpreted as WKB modes that live on in-spiral trajectories.
 For the sake of simplicity in the discussion below,
 presume that fluid elements undergo one full rotation
 in the annulus $r_{\rm ms} \geq r \geq r_{\rm mb}$ before 
 ``plunging in" from $r_{\rm mb}$.
 The integrated phase along this path can be interpreted with 
 a skewed effective wavelength which goes to increasingly larger 
 length-scales as fluid particles approach $r_{\rm mb}$.
 If we demand a tight fit of one whole, integrated WKB wavelength on the 
 in-spiral trajectory between $r_{\rm ms}$ and $r_{\rm mb}$,
 the mid Lagrangian displacement node, $\bxi(\tau) = 0$, 
 represents a converging 
 point for the flow in the linear stage of the instability 
 as seen by a co-moving geodesic observer.
 This node subsequently develops into the mass over-density 
 that we identify with a clump in the non-adiabatic stage (i.e., 
 when neutrinos have diffused out of the compressive perturbation,
 \S \ref{subs:Heat&Clump}). 
 Since the scale of the mode is larger downstream of the displacement
 node, most of the mass in the over-density will be advected in 
 from fluid particles that rush upstream toward the node 
 on the instability time scale; a smaller fraction of the clump's 
 mass flows downstream toward the node, from smaller scales upstream.

\begin{figure}
\hbox{~}
\centerline{\psfig{file=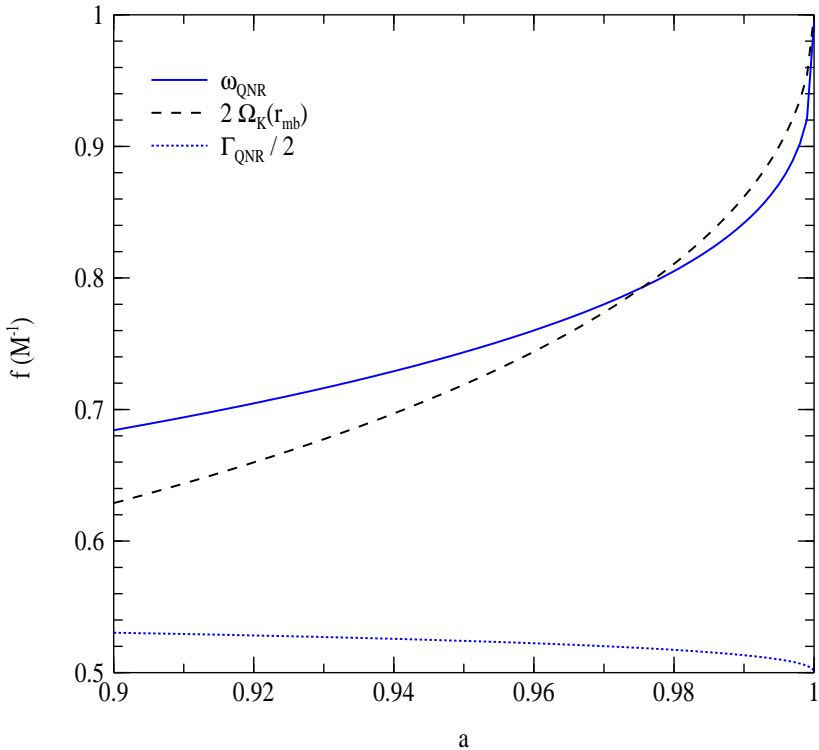,angle=0,width=5truein}}
\vskip 1mm 
\caption{
Shown are the real part of the QNR frequency, $\omega_{\rm QNR}$, 
and the driving frequency, $\omega_{\rm dr} = 2 \Omega_+ (r_{\rm mb})$,
for quadrupole oscillations
as functions of the hole's spin parameter $a$. 
Plotted also is half of the imaginary part of the QNR frequency 
offset by +0.5 from zero for ease of comparison 
with the difference among the other two frequencies.
Most notable is the fact that throughout the range 
$a \geq 0.95,
~ |\omega_{\rm QNR} - \omega_{\rm dr}| \leq \half \Gamma_{\rm QNR}$
which implies resonant driving of the QNR modes from hyper-accretion. 
}
\end{figure}

 	When the effective WKB scale on the instability is suitably large,
 a reasonably large fraction of the mass in the annulus 
 will reside in a single clump by the time it reaches $r_{\rm mb}$.
 The subsequential ``free-fall" from $r_{\rm mb}$
 (presuming this radius corresponds to the cusp in effective potential; 
 e.g., Kozlowski \etal 1978, Abramowicz \etal 1978),
 occurs on a dynamical time scale: 
 $t_{\rm dyn} \equiv \Omega_+^{-1} (r_{\rm mb})$.
 Furthermore,  
 since $l$-pole modes couple to the $l$-multipole of the dynamical frequency
 (e.g., Ex. 3.7 of Rybicky \& Lightman 1979), 
 the driving frequency for quadrupole oscillations corresponds
 to twice the relativistic Keplerian frequency at 
 $r_{\rm mb}:~ \omega_{\rm dr} \equiv 2 \Omega_+(r_{\rm mb})$. 
 Remarkably, 
 {\it the difference between the driving frequency and 
 the resonant (2,2) QNR frequency resides within a factor of 
 the order of the damping rate throughout the range
 $.9 \ltaprx a \ltaprx .99$} (Fig 3);
 whereas the Q value of the mode hovers on $5.6 < {\rm Q} < 33$.

	This picture of hyper-accretion would 
 portray an ideal scenario to drive the QNR modes resonantly 
 {\it were} 
 the clumps to arrive at $r_{\rm mb}$ steadily, 
 with a small spread around time intervals of $\cO[\Omega^{-1}_+]$,
 and for long enough to build the strain amplitude up 
 to a saturation point--or about 2Q cycles (see below).
 It is noteworthy that relativistic magneto-rotational dynamics 
 (\S \ref{sec:ISCOphys}) indicates 
 that mass over-densities form rapidly enough to feed the QNR
 modes of the hole efficiently.
 Yet, nature will add randomness to the process.

 	The added stochasticity in the arrival times of the clumps 
 is related to the seed density perturbations in the flow.
 The accretion flow outside $r_{\rm ms}$ is turbulent on small scales,
 $\ell_{\rm turb} \ll r$, and these density perturbations act as seeds
 for clump formation as these enter the relativistic annulus.
 This fact notwithstanding, 
 in the picture of clump formation outlined above 
 the probability distributions for arrival times and clump mass 
 are strongly affected by large scale magneto-rotational dynamics 
 near $r_{\rm mb}$.  
 If the flow is steady, large scale dynamics should 
 minimize the impact of small scale perturbations upstream
 {\em and} induce a correlation between the two distributions. 
 Below, we simply assume that the probability distribution for
 arrival times will favor resonant driving of the modes 
 with a moderate spread around the resonant frequency of arrival times
 (see further discussion on this point below).  
 Lastly, note that the arguments above do not depend on in-spiral 
 trajectories wrapping around fully on the annulus;  
 since the MRI time scale is suitable fast,
 a small delay to reach the free-fall stage is all that is needed 
 for MRI dynamics to act.

\begin{figure}
\hbox{~}
\centerline{\psfig{file=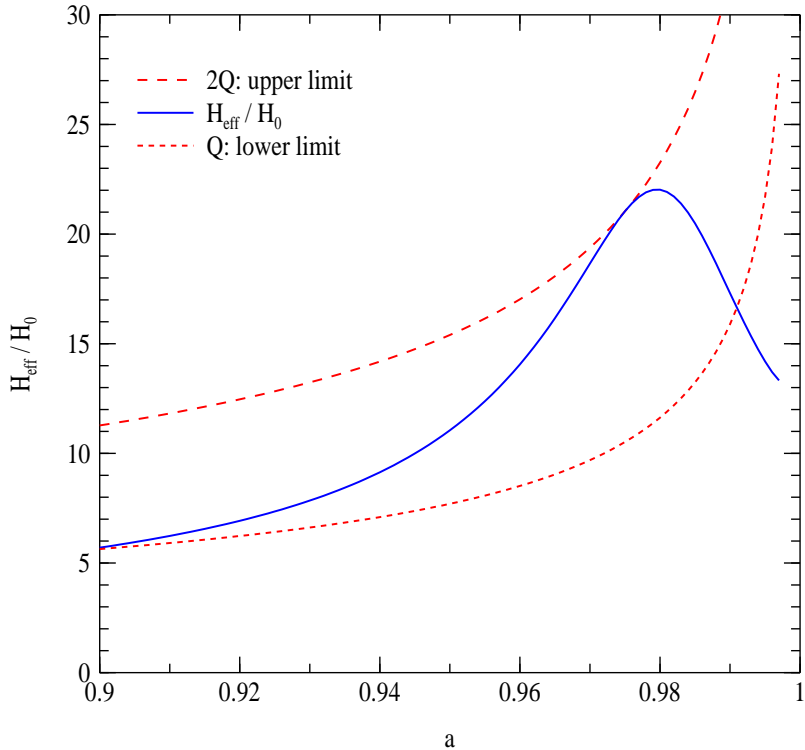,angle=0,width=5truein}}
\vskip 1mm 
\caption{
The solid line indicates the augmented amplitude of QNR oscillations
driven near resonance by clumpy hyper-accretion from a neutrino-cooled
disk as a function of the hole's spin parameter $a$. 
Modulo a small spread caused by the stochasticity in time arrival
of the clumps to $r_{\rm mb}$ (see discussion in the text),
the enhanced amplitude lies within the curves for Q($a$) and 2Q($a$)
throughout the range $a > .9$. 
A brisk energy deposition rate $\propto (H_{\rm sat}/H_0)^2$ ensues.
}
\end{figure}

	Beyond an initial transient period where the amplitude of the
 oscillation builds up to a saturation point, driven QNR oscillations
 become {\it undamped} oscillations at the driving frequency  
\beq	h(t) = {H_0 \over d} 
		\frac	{ \omega^2_{\rm QNR} } 
			{ \omega^2_{\rm dr} - \omega^2_{\rm QNR} 
			- i \, \omega_{\rm dr} \Gamma_{\rm QNR} 	} 
	~ S^2_2 (\phi,\theta,a) \, e^{-i \, \omega_{\rm dr} t} 
	\equiv 	{H_{\rm sat} \over d} ~ S_{22} (\phi,\theta,a) 
	{ e^{-i \, (\omega_{\rm dr} t + \delta_t )} },
\label{eq:DrivQNR} \eeq
 where 
 $\delta_t = \tan^{-1}(\omega_{\rm dr} \Gamma_{\rm QNR}/  
			[\omega^2_{\rm dr} - \omega^2_{\rm QNR}]) $ 
 is a slowly varying function of time
 and the saturation amplitude, $H_{\rm sat}$, represents an effective 
 strain length (in units of the system's total mass $M$) set by 
 the condition that the energy input from clump in-fall 
 ($\propto H_{\rm sat}$) be balanced by the damping losses 
 ($\propto H_{\rm sat}^2$).
 When the driving frequency is near resonance, 
 $H_{\rm sat}$ is limited primarily by the 
natural\footnote{
Quasi-normal modes are strongly damped because of the B.C.
of purely ingoing radiation at $r_+$ which saps most of the 
energy emitted by the metric perturbation unless the waves are strongly
beamed from relativistic motion (Sterl Phiney, Priv. Comm.).  
In this sense, the damping of the mode corresponds to radiation 
reaction with most of the energy being advected in by the hole. 
} damping of the QNR mode,  
 $\Gamma_{\rm QNR} \equiv \Im [\omega_{22}]$, 
 and secondarily by the relative difference of the driving frequency
 to the resonant frequency for quadrupole ringing,
 $\omega_{\rm QNR} \equiv \Re [\omega_{22}]$.
 Furthermore, 
 since $\omega_{\rm dr} \approx \omega_{\rm QNR}$ 
 for a relatively broad range of $a$ values (Fig 3),
 the resonant frequency may be taken to be $\omega_{\rm QNR}$. 

 	The energy flux carried by driven QNR waveforms benefits greatly
 from the build up in the amplitude of the strain,
 $H_{\rm sat}$, when compared to the ringing from a single excitation
 event $H_0$(c.f., Eq [\ref{eq:DampQNR}]). 
 Indeed, when driven near resonance the augmented amplitude
 is proportional to twice the Q value of the mode: 
 $H_{\rm sat}|_{\omega_{\rm dr} = \omega_{\rm QNR}} = 2\,{\rm Q}\,H_0$
 (c.f. Eq [\ref{eq:Qval}]). 
 Although the modes are driven precisely 
 at resonance only for $a \approx .98$, note from Fig 3 that
 $|\omega_{\rm QNR} - \omega_{\rm dr}| \leq \Gamma_{\rm QNR}$ 
 throughout the range $.90 \ltaprx a \ltaprx .99$. 
 This sets a lower bound on the augmented strain amplitude for this 
 range of $a$: $H_{\rm sat} / H_0 \gtaprx \, {\rm Q}$
 (recall that the resonant denominator only looks like a Lorentzian
 very near the resonance).
 Fig 4 shows the exact value of the augmented strain 
 amplitude for $a \geq .9$.
 Furthermore, adopting a very modest $\varepsilon_0 = .02$ for the 
 in-fall of a single clump (in accordance with the DRPP and LP97 results; 
 see also discussion in \S \ref{subs:SingleQNR}), the augmented efficiency
 factor in GW energy deposition from driven oscillations 
 $\varepsilon = (H_{\rm sat}/H_0)^2 \varepsilon_0$ 
 may be larger than unity.
 Depending on the hole's spin, 
 $\varepsilon \simeq 1.-10$ for $.92 \leq a \leq .98$ (c.f. Fig 4).

	Next, let us relate the energy emitted by the sequential in-fall 
 of $N$ clumps of mass $\delta M$ through $r_+$
 to the amplitude of the waveform in the time domain.  
 Assume $N$ to be small enough such that $N \delta M <  M_{\rm initial}$
 but large enough to enable a sensible average of the RMS strain 
 amplitude over the corresponding number of wave cycles.
 The energy flux averaged over N cycles is (Isaacson 1968)
\beq
\frac{d E}{dA dt} = \left[ {c^3 \over G} \right]
		{1 \over {16\pi}} 
		\langle \dot{h}(t) \cdot \dot{h}^*(t) \rangle.
\label{eq:GWflux}
\eeq 
 With $h(t)$ as given by Eq [\ref{eq:DrivQNR}] and 
 assuming no spin evolution, $\dot{a} \doteq \emptyset$,
 the total energy radiated in gravitational waves results from  
 integrating Eq [\ref{eq:GWflux}] over an encompassing surface 
 (recall $\int d\Omega |S_{22}|^2 = 1$), 
 and over a time interval corresponding to $N$ cycles of $\omega_{\rm dr}$: 
\beq
 E_{\rm GW} = 	\left[ {c^3 \over 8 G} \right] N \omega_{\rm dr} 
		\, |H_{\rm sat}|^2.
\label{eq:E-H_rel} \eeq

	On the other hand, 
 an approximate estimate of the total GW energy deposition 
 which accounts for a slight mass increment of the black hole 
 without spin evolution is as follows.
 First, write the energy emitted during one driven cycle as follows
 $\delta E = 
 [c^2] \, (\varepsilon \delta M) \, {\delta M} / M_{\rm initial}$,
 and note that the clump mass 
 $\delta M \simeq \dot{M} \Omega_+^{-1}(r_{\rm mb}) \propto M$
 (constant $a$ and $\dot{M}$).
 For $\Delta M \equiv \Sigma \, \delta M < M_{\rm initial}$,  
 replace the factor in parenthesis by the geometric mean of 
 $\delta M$ over the hyper-accreting epoch,  $\bar{\delta M}$,
 and perform the integral of $\delta M / M(t)$ by writing   
 ${\delta M} = \dot{M} dt$.  This yields
\beq
\Delta E_{\rm GW} = [c^2] \, \varepsilon \bar{\delta M} 
		\, \ln \left( {M_{\rm final} \over M_{\rm initial}} \right)
\label{eq:E_deposit}
\eeq  
 where $\varepsilon$ may be larger than unity for black hole spins 
 in excess of $a \gtaprx .92$.


\section{Detectability Criterion}

	Rather than attempt to describe the complex time evolution 
 of the gravitational waveform throughout the hyper-accreting stage, 
 we will simply assess the detectability of the waves with the matched
 filtering technique by utilizing the above energy estimate,
 Eq [\ref{eq:E_deposit}], while focusing on a relatively 
 short time interval such that the added mass to the 
 hole is a small fraction of the initial mass.  

 	For definiteness, we adopt 
 $M_{\rm initial} = 15 \msun$ (f $\simeq 2147$ Hz)
 and $\Delta \, M = 1 \msun$ corresponding to a second of hyper-accretion at 
 $\dot{M} = 1 \msun \sec^{-1}$.
 Under the simplified model for clump mass probability distribution 
 as outlined above (quasi-linear analysis), a large fraction of the 
 mass in the locus $r_{\rm mb} \leq r \leq r_{\rm ms}$ at any given time
 resides in a clump by the time fluid particles
 reach $r_{\rm mb}$.  We estimate the mass in this clump to be
 $\delta M = \dot{M} \Omega^{-1}_+(r_{\rm mb})$
 which is a factor of $2\pi$ smaller than the total 
 mass in the annulus 
 for fully wrapping, in-spiral trajectories, \S \ref{subs:ResDrivQNR}.

   	We follow Flanagan \& Huges (1998) to evaluate the 
 likelihood of detection of GW from driven QNR modes by the LIGO
 interferometers.
 The criterion for detectability of the wave train   
 through the matched filtering technique is a non-trivial S/N ratio 
\beq
	\langle [{\rm S/N}]^2 \rangle = \int d \, {\ln \omega} \, 
	\frac {h^2_{\rm char}(\omega)} {h^2_{\rm noise}(\omega)} 	
\label{eq:SN_ratio} \eeq
 where $h_{\rm noise}(\omega) ~(= h_{\rm noise}(f)/\sqrt{2\pi})$ 
 reflects the detector spectral density of 
 strain noise appropriately averaged for random incident orientations
 with respect to the LIGO interferometers.  
 The characteristic strain amplitude of the wave train is given by 
\beq
 	h^2_{\rm char} (\omega) 
	\equiv \frac {2(1+z)^2} {\pi^2 d^2} 
	{d E \over {d \omega}} [(1+z)\omega], 
\eeq
 where $d$ is the luminosity distance to the source, $z$ the redshift,
 and $\omega$ the observed angular frequency.
 Defined this way, $h_{\rm char}$ is not trivially related  
 to the amplitude of the strain in the time domain, $H_{\rm sat}$, 
 nor to the same in the frequency domain.  
 It is rather a gauge of the time integrated power spectrum weighted 
 by the ``mono-chromaticity" of the wave train over the observation time.  

 	With $(1+z)\omega \rightarrow \omega$,
 the GW energy spectrum from driven QNR modes in the radiation zone,
\beq
	{d E(\omega) \over {d \omega}} 
	= \frac {\omega^2}{16 \pi^2} 
	\, \tilde{H}(\omega) \cdot \tilde{H}^*(\omega),
\eeq
 ensues from a Fourier transform of the strain
 signal in the time domain as given in Eq [\ref{eq:DrivQNR}]
 (compare this expression with Eq [2.38] of Flanagan \& Huges 1998,  
 and note the conforming notation used throughout).

	Let us approximate all quantities $\propto M$ by their geometric
 mean and assume no spin evolution during a small epoch of hyper-accretion
 $T = 2 \pi N / \omega_{\rm dr}$ (with $\log N$ in the range 3-3.5).
 With $\tilde{H} (\omega) = 2 {H_{\rm sat} e^{i\delta_t}} 
	\sin [{{N\pi \Delta \omega}/ \omega_{\rm dr}}] / \Delta \omega$
 (note that $\dot{a} \doteq 0 \Rightarrow \delta_t = $ constant as well),
 the characteristic strain amplitude of driven QNR waves 
 turns out to be  
\beq
 	h^2_{\rm char} (\omega) 
	\equiv \frac {8(1+z)^2} {(2\pi)^4} 
	{{H_{\rm sat} \cdot H^*_{\rm sat}} \over {d^2} } \,
	\left[ { N \pi \omega \over \omega_{\rm dr} } \,
	\frac {\sin (x-b)}{x-b} 
	\right]^2
\eeq
 where $x-b \equiv N \pi (\omega - \omega_{\rm dr})/ \omega_{\rm dr}$.
 
 	The strength of the characteristic strain signal at the driving 
 frequency $\omega = 2\Omega_+(r_{\rm mb})$ follows by setting 
 the expression in the square parenthesis to $[N \pi]^2$
 and by utilizing Eq [\ref{eq:E-H_rel}] 
 to replace $|H_{\rm sat}|^2$ in favor of the energy released 
 during the hyper-accreting epoch $\Delta E_{\rm GW}$ as given by 
 Eq [\ref{eq:E_deposit}]
\beq
 	\left. h^2_{\rm char}
	\right|_{\omega = \omega_{\rm dr}} 
	= \left[ G \over c^3\right]
	  \left( {2 (1+z) \over \pi d} \right)^2 
	  { N \Delta E_{\rm GW} \over {\omega_{\rm dr}} }
	= \left[ G \over {2 \pi c} \right] 
	\left( {2 (1+z) \over \pi d} \right)^2 
  	\, \varepsilon \bar{\delta M} \, T \, 
	\ln \left( {M_{\rm final} \over M_{\rm initial}} \right).
\eeq

	For GRB 030329, we plug in $z=.1685$, $d= 810$Mpc
 ($H_0 = 70, ~\Omega_{\rm b} = .3, ~\Omega_{\rm v} = .7$),
 $M_{\rm initial} = 15 \msun$, $T = 1 \sec$, and  
 $ \bar{\delta M} =  \dot{M} \Omega_+^{-1} (r_{\rm mb})
 \simeq 1.83_{-4} \msun$
 and choose $a= .98$ ($H_{\rm sat}/H_0 = 22$) 
 to obtain
 $h_{\rm char} \simeq 8.4_{-23}$ 
 at the observed linear frequency $f = 1490{\rm Hz}$.
 Utilizing similar parameters for GRB980425 at $d= 27$Mpc, 
 turns out a characteristic signal
 $h_{\rm char} \simeq 2.16_{-21}$ 
 at $f \simeq 1741$Hz.

\section{Discussion}

	We have laid out a physical--if simple minded--model for resonant
 driving of the quasi-normal ringing (QNR) wave modes of the Kerr geometry.
 A nascent black hole hyper-accreting at rates 
 $\dot{\rm M} \approx 1 \msun \sec^{-1}$ from a neutrino cooled disk 
 is reckoned to oscillate near resonance of its ($l,m= 2,2$) quadrupole QNR 
 frequency due to the in-fall of compact mass over-densities from the cusp 
 in effective potential on a dynamical time scale.
 This model is based on large-scale, magneto-rotationally--induced fluid
 dynamics in the ultra-relativistic region of the flow bounded from below
 by the marginally bound orbit radius: 
 $r_{\rm mb}$
 (assumed to coincide with the aforementioned cusp). 
 The exact location of the cusp will change 
 the dynamical time scale for in-fall but not by much when the spin 
 parameter $a \geq .9$.
 Furhermore,
 since the MRI time scale is suitable fast, a small delay to reach the 
 free-fall stage is all that is needed for MRI dynamics to act.
 Heat conduction from neutrino diffusion out of mass over-densities
 on the time scale of the instability 
 warrants that compressive perturbations behave non-adiabatically.
 This leads to sufficiently long-lived clumps in 
 a highly intermittent medium.
 When the spin parameter of the hole exceeds 90\%,
 large-scale magneto-rotational wave modes will segregate
 compact matter over-densities from large-scale magnetic field domains.  
 The subsequent in-fall of these clumps from $r_{\rm mb}$
 will drive the QNR modes of the geometry in resonant fashion
 if the clumps arrive to $r_{\rm mb}$ steadily, 
 with a small spread around time intervals of $\cO[\Omega^{-1}_+]$,
 and for long enough to build the strain amplitude up 
 to a saturation point.
 
 It is difficult to address the probability distribution for 
 clump arrival times from the linear theory alone but one anticipates
 that it will involve a co-moving time scale of $\cO[\Omega^{-1}_+]$
 and the integrated, spin dependent WKB wavelength on the inspire 
 trajectory toward $r_{\rm mb}$.
 We have argued that for large mass clumps,
 $\delta M \simeq \dot{M}\Omega_+^{-1}$, 
 the arrival times to $r_{\rm mb}$ will not be entirely stochastic 
 since ``the bucket needs time to fill-up", nor entirely coherent 
 since the seed density perturbations are stochastic but large-scale
 MRI dynamics will work to minimize the impact of this initial
 small-scale randomness.
 If the arrival times were entirely stochastic, 
 the enhancement in energy deposition would be roughly proportional
 to the Q value of the mode instead of to Q$^2$ 
 and the corresponding signal would be smaller by about an order of 
 magnitude.
 A detailed probability distribution of arrival times requires an 
 accurate account of the radial velocity profile and of the 
 effective, spin-dependent WKB wavelengths on in-spiral trajectories
 which is beyond the scope of this study.

	If the QNR modes are fed resonantly for a few seconds
 of hyper-accretion, the enhanced amplitude of the oscillations
 yields a very high rate of energy deposition into gravitational waves. 
 Indeed, the integrated energy deposition is large enough to ``evaporate"
 the equivalent of a factor of a few times the total 
 rest-mass--energy of a single clump into gravitational waves,
 Yet, the characteristic frequency of emission resides on 
 a less than optimal range with respect to the 
 advanced LIGO noise curve: $f\simeq 8000-1600$Hz for $M \simeq 3-15 \msun$.
 On the other hand, 
 the high quality and coherence of the expected signal 
 should make this model a prime candidate for narrow-band LIGO searches.
 For gamma-ray bursts at low redshifts, 
 this may ultimately lead to a signal strong enough to be seen by LIGO II.

	A 15$\msun$ black hole accreting at $1 \msun \sec^{-1}$
 may come about from the death throes of a very massive star.
 If such an explosion is associated with a long GRB,
 and the broad properties of the prompt gamma-ray light curve, 
 to inner engine activity;
 hyper-accretion onto this heavy weight, fast spinning hole
 should be associated with a second broad hump in the light curve.
 We anticipate that gravitational waves with the predicted properties
 may be present at this stage of the GRB. 
 The proposed mechanism is thus a direct probe of the elusive 
 inner engine of gamma-ray bursts.

\vfill
\section*{Acknowledgments}

 It is a great pleasure to acknowledge stimulating and instructive 
 discussions with Sterl Phinney, Lee Lindblom, Kip Thorne, Ethan Vishniac 
 and Roger Blandford.  I am greatly indebted with the TAPIR Group at Caltech 
 for their hospitality and to Don Juan's teachings;
 {\it ``...because the warrior, having chosen a path, has but one goal:
 to traverse its full length."}


{\footnotesize


\begin{thebibliography}{99}

\bibitem{Abram+78} 					
 	Abramowicz, M., Jaroszynski, M. \& Sikora, M. 1978,
	A\&A, 63, 221

\bibitem{AgoKro98} 					
	Agol, E. \& Krolik, J. 1998, Ap.J., 507, 304. 

\bibitem{AG02} 				
	Araya-G\'ochez, R.A. 2002, M.N.R.A.S. 337, 795


\bibitem{AGVis02} 				
 	\AG, R.A. \& Vishniac, E.T. 2002, M.N.R.A.S. submitted
	(astro-ph/0208007)

\bibitem{BH91} 				
 	Balbus, S.A. \& Hawley, J.F. 1991, Ap.J. 376, 214

\bibitem{BH92a} 				
	Balbus, S.A. \& Hawley, J.F. 1992, Ap.J. 392, 662 

\bibitem{BH98} 				
	Balbus, S.A. \& Hawley, J.F. 1998, Rev. Mod. Phys. 70, 1 


\bibitem{BaPrTe72}
 	Bardeen, J., Press, W. \& Teukolsky, S. 1972, Ap.J., 178, 347 

\bibitem{BlaSoc01}
 	Blaes, O. \& Socrates, A. 2001, Ap.J., 553, 987 				
\bibitem{BlaBeg99} 					
 	Blandford R. \& Begelman M. 1999, M.N.R.A.S., 303, L1 

\bibitem{BlaBeg03} 					
 	Blandford R. \& Begelman M. 2003 
	M.N.R.A.S. submitted (astro-ph/0306184)

\bibitem{CurPud95} 				
	Curry, C. \& Pudritz, R. 1995, Ap.J., 453, 697 	

\bibitem{DRPP71} 					
 	Davis, M., Ruffini, R., Press, W. \& Price, R. 1971,
 	Phys. Rev. Lett., 27, 1466

\bibitem{DiMatt+02} 					
 	Di Matteo, T., Perna, R. \& Narayan, R. 2002, Ap.J., 579, 706 

\bibitem{Echev89} 					
	Echeverria, F. 1989, Phys. Rev. D, 40, 3194	

\bibitem{FlaHug98} 					
 	Flanagan, E. \& Huges, S. 1998, Phys. Rev. D, 57, 4535

\bibitem{Fogli95} 					
 	Foglizzo, T. 1995, Ph.D. Thesis, University of Paris VII 

\bibitem{FT95} 					
 	Foglizzo, T. \& Tagger M. 1995, A.\&A., 301, 293 

\bibitem{Fryer+02}
 	Fryer, C., Holz, D. \& Hughes, S. 2002, Ap.J., 565, 430 
	(astro-ph/0106113)		 		
 
\bibitem{GamPop98}
	Gammie, C.F. \& Popham, R. 1998, Ap.J., 498, 313 					
\bibitem{Thorne87} 				
	K.S Thorne 1987, in 
 	{\it 300 Years of Gravitation} \\
 	eds: S. Hawking \& R. Israel  
 	Cambridge Univ. Press

\bibitem{HawBal02} 				
	Hawley, J.F. \& Balbus, J.H 2002, Ap.J., 573, 738

\bibitem{HawKro01} 				
	Hawley, J.F. \& Krolik, J.H 2001, Ap.J., 548, 348 

\bibitem{HawKro02} 				
	Hawley, J.F. \& Krolik, J.H 2002, Ap.J., 566, 164 
	 \& Ap.J. 573, 754 

\bibitem{Isaac68}
 	Isaacson, R.A. 1968, Phys. Rev., 166, 1272

\bibitem{Kozlo+78}
 	Kozlowski, M., Jaroszy\'nski, M. \& Abramowicz, M. 1978
	A\&A, 63, 209

\bibitem{Krolik99}
	Krolik, J.H. 1999,
	{\it Active Galactic Nucleii: 
	from the Central Black Hole to the Galactic Environment} \\
	Princeton Series in Astrophysics, 
 	Eds. Jeremiah P. Ostriker \& David N. Spergel
	(Princeton: Princeton University Press)

\bibitem{Leav85}
	Leaver, E.W. 1985, Proc. R. Soc. Lond. A 402, 285 

\bibitem{LouPri97}
  	Lousto \& Price 1997, Phys. Rev. D 55, 2124L 

\bibitem{NarYi95}
	Narayan, R. \& Yi, I. 1995a, Ap.J., 444, 231; 
			  ~~1995b, Ap.J., 452, 710 

\bibitem{Naray+00}
	Narayan, R., Igumenshchev, I. \& Abramowicz, M. 2000,  
 	Ap.J., 539, 798; 

\bibitem{NovTho73} 
	Novikov, I. \& Thorne, K. 1973, in {\it Black Holes} \\
	eds. C. DeWitt \& B.S. DeWitt (New York: Gordon \& Breach), 343

\bibitem{OgiPri96}
 	Ogilvie, G.I. \& Pringle, J.E. 1996, M.N.R.A.S., 279, 152

\bibitem{PopGam98}
 	Popham, R. \& Gammie, C.F. 1998, Ap.J., 504, 419 					
\bibitem{PWF99}
 	Popham, R., Woosley, S., \& Fryer, C. 1999, Ap.J., 518, 356

\bibitem{RybLig79}
  	Rybicky, G. \& Lightman, A. 1979 
	{\it Radiative Processes In Astrophysics} \\
 	John Wiley \& Sons, Inc

\bibitem{ShaSun}
	Shakura, N.I. \& Sunyaev, R.A. 1973, A\&A, 24, 337					
\bibitem{Teuko73}
 	Teukolsky, S. 1973, Ap.J., 185, 635 

\bibitem{Turn+01}
	Turner, N., Stone, J., \& Sano, T. 2001 (astro-ph/0110272) 				
\bibitem{Zeril70} 				
 	Zerilli 1970, Phys. Rev. D,  2, 2141 

\end{thebibliography}
\end{document}